\documentclass[aps,preprint,preprintnumbers,superscriptaddress,nofootinbib,]{revtex4-1} 

\usepackage{amsmath,amssymb}    
\usepackage[dvipdfmx]{graphicx}   
\usepackage{verbatim}   
\usepackage{color}      
\usepackage{subfigure}  
\usepackage{hyperref}   
\usepackage{here}       
\usepackage{bm}         

\begin{document}

\preprint{KOBE-COSMO-19-04, RUP-19-7}

\title{Dressed Power-law Inflation with Cuscuton}

\author{Asuka Ito}
\email[]{asuka-ito@stu.kobe-u.ac.jp}
\affiliation{Department of Physics, Kobe University, Kobe 657-8501, Japan}

\author{Aya Iyonaga}
\email[]{iyonaga@rikkyo.ac.jp}
\affiliation{Department of Physics, Rikkyo University, Toshima, Tokyo 171-8501, Japan}

\author{Suro Kim}
\email[]{s-kim@stu.kobe-u.ac.jp}
\affiliation{Department of Physics, Kobe University, Kobe 657-8501, Japan}

\author{Jiro Soda}
\email[]{jiro@phys.sci.kobe-u.ac.jp}
\affiliation{Department of Physics, Kobe University, Kobe 657-8501, Japan}

\date{\today}

\begin{abstract}
We study dressed inflation with a cuscuton and find a novel exact power-law solution.
It is well known that the conventional power-law inflation is inconsistent with the Planck data.
In contrast to this standard lore, we find that power-law inflation with a cuscuton can be reconciled with the Planck data. 
Moreover, we argue that the cuscuton generally ameliorates inflation models so that predictions are consistent with observations. 
\end{abstract}

\maketitle


\section{Introduction}
An inflationary scenario provides a mechanism for generating temperature fluctuations 
of cosmic microwave background  radiations (CMB) and the large scale structure of the universe.
It is always true that an exact solution is useful for obtaining a profound understanding 
of a physical mechanism.
In the case of inflationary scenarios, power-law inflation is known as an exact solution~\cite{Lucchin:1984yf}.
Observationally, however, recent precision data have ruled out the power-law inflation~\cite{Akrami:2018odb}. The reason is as follows.
The key relations of the power-law inflation
 are given by
\begin{eqnarray}
 n_s -1 = -2 \epsilon \ , \quad
 r= 16\epsilon  \ ,
\end{eqnarray}
where  $n_s$, $r$, and $\epsilon$ are the scalar spectral index, the tensor-to-scalar ratio, and   a slow-roll parameter, respectively.
Since the CMB data tells us $n_s \sim 0.96$, we obtain the tensor-to-scalar ratio
\begin{eqnarray}
 r=  8(1-n_s) \sim 0.3
\end{eqnarray}
Apparently, this contradicts the Planck constraint $r\leq 0.1$~\cite{Akrami:2018odb}.
Thus, the power-law inflation is a failed inflation model.

The question we raise in this paper is if we can ameliorate a failed inflation model so that the predictions are consistent with observational data. To settle this issue, we utilize an exact solution, namely, power-law inflation. 
 We focus on the minimal models of inflation
which include two degrees of freedom for gravity and one for inflaton. 
 This type of models can be conventionally described by Einstein gravity with an inflaton field. Intriguingly, the minimal theory can be extended nontrivially if we include a non-dynamical scalar field called cuscuton~\cite{Afshordi:2006ad}. 
 The gravity with cuscutons, which we call cuscuton gravity
 in this paper, is the infrared modification of gravity.
 In fact, the application of cuscuton gravity to the late time accelerating universe is discussed in \cite{Afshordi:2007yx}. The systematic analysis of perturbations has been also performed~\cite{Boruah:2017tvg}.
It turns out that cuscuton gravity is related to the low energy limit of
Lorentz violating gravity~\cite{Afshordi:2009tt, Bhattacharyya:2016mah}.
The subtle point of cuscuton gravity is also analyzed from the perspective of the Hamiltonian analysis~\cite{Gomes:2017tzd}. 
It is shown that the cuscuton has to be homogeneous to be non-dynamical.
Moreover, it is shown that cuscuton gravity can be extended to
 the more general theories~\cite{Iyonaga:2018vnu}.
The cuscuton gravity is  relevant not only to the late time cosmology
 but also to the early universe. 
Indeed, the cuscuton allows us to make the bounce universe consistent~\cite{Boruah:2018pvq}.

In this paper, we will consider inflation in the context of cuscuton gravity. 
We  find a new exact power-law solution which can be regarded as dressed power-law
 inflation with a cuscuton. 
Although the conventional power-law inflation is ruled out by the Planck data~\cite{Akrami:2018odb}, we show that
the power-law inflation can be reconciled with 
the CMB data in the presence of the cuscuton.
To show this, we study equations for curvature perturbations and tensor perturbations. 
Since the kinetic term is non-trivial in cuscuton gravity, we need to care about
 the gradient instability. In contrast to the usual gradient instability~\cite{Kawai:1998ab,Kawai:1999pw}, the gradient instability in cuscuton gravity could exist in the infrared regime even when it exists. 
We explicitly show there is no instability in the dressed power-law inflation.
Moreover, we calculate the scalar spectral index and the tensor-to-scalar ratio
and find that the dressed power-law inflation can be reconciled  with observations.
This provides a positive answer to the issue we raised.

The paper is organized as follows: In section II, we introduce cuscuton gravity and
present exact solutions for power-law inflation with a cuscuton.
In section III, we calculate the spectral index and the tensor-to-scalar ratio. We show power-law inflation with a cuscuton is consistent with the CMB data. The final section is devoted to the conclusion.

\section{Power law inflation in cuscuton gravity}
We consider a dressed inflationary universe driven 
by an inflaton field $\chi(x)$ with a cuscuton field 
$\phi(x)$~\cite{Afshordi:2006ad} .
The action for cuscuton gravity is given by 
\begin{eqnarray}
  S =  \int d^{4}x \sqrt{-g} 
      \Bigg[ \frac{M_{pl}^{2}}{2}R \pm \mu^{2}  \sqrt{- \partial_{\mu}\phi \partial^{\mu}\phi} -V(\phi) 
           -\frac{1}{2} \partial_{\mu}\chi \partial^{\mu}\chi -U(\chi)  \Bigg] \ , \label{act}
\end{eqnarray}
where $g$ is the determinant of the metric $g_{\mu\nu}$, $M_{pl}$ is the reduced Planck mass, 
$R$ is the Ricci scalar and $\mu$ is a constant.
Here, $V(\phi)$ and $U(\chi)$ are potentials for the cuscuton and the inflaton,
respectively.
For the background spacetime, we take the flat FRW metric as
\begin{equation}
  ds^{2} = -dt^{2} + a(t)^{2} dx_{i}dx^{i} \ ,  \label{met}
\end{equation}
where $a(t)$ is the scale factor. 
From the action (\ref{act}) and  the metric (\ref{met}), assuming the homogeneity of the cuscuton and the inflaton,
one can derive the hamiltonian constraint:
\begin{equation}
  H^{2} = \frac{1}{3M_{pl}^{2}} \left[ V + U + \frac{1}{2} \dot{\chi}^{2} \right] \ , \label{hami}
\end{equation}
the Einstein equation:
\begin{equation}
    \dot{H} = -\frac{1}{2M_{pl}^{2}} \left[ \pm \mu^{2} |\dot{\phi}| +  \dot{\chi}^{2}  \right] \ , \label{ein}
\end{equation}
and the field equations:
\begin{eqnarray}
  \pm {\rm sign}(\dot{\phi}) 3\mu^{2} H + V_{,\phi} &=& 0 \ , \label{eom1} \\ 
  \ddot{\chi} + 3H\dot{\chi} + U_{,\chi} &=& 0 \ .  \label{eom2}
\end{eqnarray}
Here $H \equiv \dot{a}/{a}$ is the Hubble parameter.
Eq.\,(\ref{eom1}) shows that the cuscuton is non-dynamical because 
a second derivative term is absent.

In the conventional Einstein gravity, we know there exists power-law solutions for
 the exponential potential
\begin{eqnarray}
  U(\chi) = U_{0} e^{u\frac{\chi}{M_{pl}}} \ . \label{exp}
\end{eqnarray}
 In the present case, to have scaling solutions, we need to take the quadratic potential for the cuscuton
\begin{eqnarray}
  V(\phi) = \frac{1}{2}m^{2}\phi(t)^{2} \ . \label{quad}
\end{eqnarray}
Let us seek for an exact solution of Eqs.\,(\ref{hami})-(\ref{eom2}) by taking the following ansatz:
\begin{eqnarray}
  H(t) = \frac{p}{t}, \quad \frac{\chi(t)}{M_{pl}} = s \ln M_{pl} t
  , \quad \phi(t) = \frac{q}{t}   \ . \label{ans}
\end{eqnarray}
Note that there is a freedom of a constant shift of $\chi$ which can be
absorbed into $U_0$.
For the quadratic potential for the cuscuton, one can take the branch
 $\dot{\phi} > 0$ without loss of generality.
Then, from Eq.\,(\ref{eom1}), we see that the plus sign of the first term must be chosen to realize an expanding universe.

Substituting the ansatz \,(\ref{ans}) into Eqs.\,(\ref{hami})-(\ref{eom2}), we can 
find sets
of exact inflationary solutions. First, we notice that the following solutions  
\begin{equation}
  m=0, \quad \mu=0, \quad 
  u=-\frac{2}{s}, \quad 
  \frac{U_{0}}{M_{pl}^{4}} = \frac{3}{4}s^{4} -\frac{1}{2}s^{2}, \quad 
  p= \frac{1}{2}s^{2} \ ,
  \label{power}
\end{equation}
 correspond to the original power-law inflation~\cite{Lucchin:1984yf} 
in the absence of  the cuscuton.
It is easy to see that  sufficient inflation occurs if $p \gg 1$.
However, as we have already mentioned,
 the solution (\ref{power}) is not consistent with the observations of CMB~\cite{Akrami:2018odb}.
In cuscuton gravity, we found a new set of solutions
\begin{equation}
  u=-\frac{2}{s},  \quad 
  p = s^{2} \left( 2 - \frac{3\mu^4}{M_{pl}^2  m^2} \right)^{-1} , \quad 
  q = -\frac{3\mu^{2}s^{2}}{m^{2}} \left( 2 - \frac{3\mu^4}{M_{pl}^2  m^2} \right)^{-1} ,   \label{new}
\end{equation}
where $2 - \frac{3\mu^4}{M_{pl}^2  m^2} > 0$ should be satisfied to keep $p$ positive. We also have a relation
\begin{equation}
  \frac{U_{0}}{M_{pl}^{4}}
   = \frac{s^{2}}{2} \left( 
    \frac{3s^{2}}{  2 - \frac{3\mu^{4}}{M_{pl}^{2}m^{2}} } -1 \right) \ .
\end{equation}
This solution represents a new power-law inflationary solution modified by the cuscuton 
when $p \gg 1$.
Indeed, the slow roll parameter 
\begin{equation}
  \epsilon = - \frac{\dot{H}}{H^{2}} = \frac{1}{p} \ , \label{fir}
\end{equation}
is small when $p$ is large.
We mention that the other slow roll parameter 
\begin{equation}
  \eta = \frac{\dot{\epsilon}}{H\epsilon} \ , \label{sec}
\end{equation}
vanishes for the power-law solution.

In the next section, we calculate the scalar spectral index and 
the tensor-to-scalar ratio and show that the dressed power-law inflation with a cuscuton can be reconciled with observations.
\section{Primordial fluctuations in cuscuton gravity}
In this section, we evaluate the power spectrum of curvature perturbations and  gravitational waves in the background (\ref{new}).
It turns out that the tensor-to-scalar ratio can be tuned freely by virtue of the cuscuton,
without changing the tilt of the scalar power spectrum.  
Moreover, we check that there is no ghost and 
gradient instabilities.

To investigate the evolution and the stability of the fluctuations, 
we need the quadratic action for perturbations.
We take an uniform field gauge $\delta \chi=0$ to eliminate the perturbation of the inflaton field.%
\footnote{We confirmed that the result in the uniform field gauge is same as that in the unitary gauge,
i.e., $\delta\phi = 0$, where cuscuton gravity is well-defined~\cite{Gomes:2017tzd}. 
}
It should be mentioned that the cuscuton field $\phi$ is not dynamical and thus the perturbation of the 
cuscuton $\delta\phi$ can be removed from the second order action eventually.

In the uniform field gauge, the metric perturbations are represented by
\begin{eqnarray} 
  ds^{2} = a(\tau)^{2} \big[ -\left(1+ 2\Phi(x)\right)d\tau^{2} 
  + 2\partial_{i}B(x) d\tau dx^{i}  
  + \left( 1+ 2\zeta(x) \right) \left( \delta_{ij} +  h_{ij}(x) \right) dx^{i}dx^{j} \big]  \ ,
\end{eqnarray}
where $\tau$ is the conformal time, $\Phi$ and $\zeta$ represent
the Newton potential and the curvature perturbations, respectively.
 A transverse traceless  tensor $h_{ij}$ describes gravitational waves. 
Note that we do not need to consider vector perturbations.
Since $\Phi$, $B$ and $\delta\phi$ are non-dynamical, we can eliminate them from the action.
Thus, as is shown in \cite{Boruah:2017tvg}, 
we can obtain the second order action for the Fourier coefficient of the curvature perturbations as
\begin{equation}
 \int d\tau d^{3}k  \frac{z^{2}}{2} \Big[ \zeta'^{2}_{k} - c_{s}^{2} k^{2} \zeta_{k}^{2}  \Big] \ ,  \label{zeta}
\end{equation}
where we defined functions
\begin{eqnarray}
  z^{2} &=& 2 a^{2} \alpha \left( \frac{k^{2} + 3\alpha\mathcal{H}^{2}}{ k^{2} + \alpha(3-\sigma)\mathcal{H}^{2}}  \right) 
            \ , \\
  c_{s}^{2} &=& \frac{ k^{4} + \left( \alpha (6- \sigma)  + 2\sigma (3  -\epsilon)  \right) \mathcal{H}^{2} k^{2}
   + \left( 3\alpha^{2} (3 - \sigma )  + 4\alpha \sigma (3 - \sigma ) \right) \mathcal{H}^{4}  }
                     { k^{4} + \alpha (6- \sigma) \mathcal{H}^{2} k^{2} +  3\alpha^{2} (3 - \sigma )  \mathcal{H}^{4} } \ , 
\end{eqnarray}
and the comoving Hubble parameter $\mathcal{H} = \frac{a'}{a}$.
We note that the slow roll parameter in terms of the inflaton 
\begin{eqnarray}
  \alpha = \frac{\chi'^{2}}{2 M_{pl}^{2} \mathcal{H}^{2}} 
\end{eqnarray}
is different from that in terms of the Hubble parameter defined by Eq.\,(\ref{fir}). The difference is described by a new parameter
\begin{eqnarray}
  \sigma = \epsilon - \alpha   \ .
\end{eqnarray}
In the case of the present new power-law solution, we can express them as
\begin{eqnarray}
  \alpha 
  = \frac{1}{2s^2}\left( 2 - \frac{3\mu^4}{M_{pl}^2  m^2} \right)^2  \ ,  \quad
  \sigma 
  =\frac{3 \mu^4}{2M_{pl}^2  m^2s^2}\left( 2 - \frac{3\mu^4}{M_{pl}^2  m^2} \right)   \ .
\end{eqnarray}
We see that 
$\sigma \rightarrow 0$ in the limit $\mu \rightarrow 0$ and then 
$\alpha = \epsilon$, which is nothing but the case of conventional
single field inflation.
Therefore $\sigma$ represents contribution of the cuscuton field to inflation
and/or deviation from conventional single field inflation.

Furthermore, from Eq.\,(\ref{ein}), $\alpha$ and $\sigma$ must be order of the slow roll parameters to realize inflation.
Note that only the plus sign in Eq.\,(\ref{ein}) describes an expanding universe.
Hence, the weak energy condition is always satisfied,
which implies the stability of the system.
In fact, since $\alpha>0$ and $\sigma>0$ hold, both of $z^{2}$ and $c_{s}^{2}$ are always positive, so that there is no 
ghost and gradient instabilities.

Near the horizon crossing, Eq.\,(\ref{zeta}) reduces to that of 
conventional single field inflation, i.e., $z\simeq 2a^{2}\alpha$ and $c_{s}\simeq 1$.
We then have the power spectrum of the curvature perturbations as 
\begin{equation}
  P_{\zeta} = \frac{H^{2}}{8\pi^{2}M_{pl}^{2}\alpha}\Big|_{aH=k} \ , \label{scpo}
\end{equation}
where the right hand side is evaluated at the crossing time $aH=k$.
One can obtain the tilt of the power spectrum as
\begin{equation}
  n_{s} - 1 = \frac{d \ln P_{\zeta}}{d \ln k} \simeq -2\epsilon  \ . 
\end{equation}
The tilt is determined by the slow roll parameter in terms of the Hubble parameter.

On the other hand, the tensor perturbations are not mixed with the scalar perturbations at the linear order,
so that gravitational waves are not affected by the cuscuton.
Therefore, the power spectrum of gravitational waves is the same as that of
conventional single field inflation,
\begin{equation}
  P_{h} = \frac{2H^{2}}{\pi^{2}M_{pl}^{2}}\Big|_{aH=k} \ .
\end{equation}
Then the tensor-to-scalar ratio is given by
\begin{equation}
  r = \frac{P_{h}}{P_{\zeta}} = 16 \alpha \ .  \label{ttos}
\end{equation}
Here, it should be stressed that the slow roll parameter in terms of the inflaton $\alpha$ appears instead of $\epsilon$.

Let us express observables for dressed inflation 
using the exact background solution (\ref{new}).
Given the power-law solution, one can calculate $n_{s}$ and $r$ as 
\begin{equation}
  n_{s} -1 = -\frac{2}{s^{2}} \left( 2 - \frac{3\mu^{4}}{M_{pl}^{2}m^{2}} \right) , \quad 
  r = \frac{8}{s^{2}} \left( 2 - \frac{3\mu^{4}}{M_{pl}^{2}m^{2}} \right)^{2} .  \label{para}
\end{equation}
Note that only the red-tilt is allowed since the inside of the parenthesis is positive.
The relations in (\ref{para})  give rise to 
\begin{equation}
  r = 2 s^{2} (n_{s} - 1)^{2} \ .
\end{equation}
Thus, it turns out that the tensor-to-scalar ratio $r$ can be tuned to a small value
with the tilt $n_{s}$ fixed.
Therefore, the dressed power-law inflation in cuscuton gravity is 
consistent with the Planck data
in contrast to the original power-law inflation \cite{Lyth:1991bc}. 
In fact, when we use observed value $n_s \sim 0.96$, we have $ r = 3\times 10^{-3} s^{2} $.
The current constraint $r\leq 0.1$ gives rise to an inequality $s^2\leq 30$.

Finally, although we focused on the dressed power-law inflation as an illustration,
we believe the result would be general in any dressed inflationary scenario.
This is because the parameter $\sigma>0$  allows us to control $n_s$ and $r$ independently.
\section{Conclusion}
We studied an inflationary universe in the context of cuscuton gravity.
We found a new exact power-law inflationary solution which
is dressed power-law inflation with a cuscuton.
Furthermore, we investigated primordial fluctuations.
It turned out that there is no ghost and gradient instabilities in dressed inflation.
We calculated the scalar and the tensor power spectrum and showed 
 that 
the tensor-to-scalar ratio can be tuned freely regardless of the tilt in virtue of the cuscuton.
It implies that the dressed power-law inflation can be always reconciled with  observations.

As an extension of the present analysis, we can consider anisotropic power-law 
inflation~\cite{Kanno:2010nr,Yamamoto:2012tq,Ohashi:2013pca,Ito:2015sxj},
constant-roll inflation~\cite{Motohashi:2014ppa}, and the mixed one~\cite{Ito:2017bnn}.
Although we have studied only the dressed power-law inflation in this paper, we expect
the qualitative result holds for general inflation models.
It would be interesting to study general dressed inflation models in detail.  
It is also intriguing to investigate higher order correlations such as the non-gaussianity.
We leave these issues for future work.

\begin{acknowledgments}
A.\,Ito was supported by Grant-in-Aid for JSPS Research Fellow and JSPS KAKENHI Grant No.JP17J00216. A.\,Iyonaga was supported by the Rikkyo University Special Fund for Research.
J.\,S. was in part supported by JSPS KAKENHI Grant Numbers JP17H02894, JP17K18778, JP15H05895, JP17H06359, JP18H04589. J.S and S.K. are also supported by JSPS Bilateral Joint Research Projects (JSPS-NRF 
collaboration) String Axion Cosmology. 

\end{acknowledgments}
\end{document}